# Bio-inspired Flexible Twisting Wings Increase Lift and Efficiency of a Flapping Wing Micro Air Vehicle

David Colmenares[1,2], Randall Kania[1], Wang Zhang[1], and Metin Sitti[1,2]

[1] Department of Mechanical Engineering, Carnegie Mellon University, Pittsburgh, PA 15213, USA
**E-mail:** [dcolmena, rkania, wangz]@andrew.cmu.edu

[2] Physical Intelligence Department, Max Planck Institute for Intelligent Systems, 70569 Stuttgart, Germany
**E-mail:** sitti@is.mpg.de

**Abstract**
We investigate the effect of wing twist flexibility on lift and efficiency of a flapping-wing micro air vehicle capable of liftoff. Wings used previously were chosen to be fully rigid due to modeling and fabrication constraints. However, biological wings are highly flexible and other micro air vehicles have successfully utilized flexible wing structures for specialized tasks. The goal of our study is to determine if dynamic twisting of flexible wings can increase overall aerodynamic lift and efficiency. A flexible twisting wing design was found to increase aerodynamic efficiency by 41.3%, translational lift production by 35.3%, and the effective lift coefficient by 63.7% compared to the rigid-wing design. These results exceed the predictions of quasi-steady blade element models, indicating the need for unsteady computational fluid dynamics simulations of twisted flapping wings.

## 1. Introduction

Bio-inspired flapping wing micro air vehicles (FWMAVs) possess many unique capabilities compared to traditional aircrafts. Their extreme maneuverability, small size, and ability to operate in confined spaces makes them well suited for applications such as search and rescue, planetary exploration, environmental monitoring, and inspection. Despite their advantages, significant challenges in actuation, power, and control must be overcome to make FWMAVs reliable platforms. In previous works, we demonstrated an underactuated, motor driven FWMAV capable of liftoff and controlling torques with a weight of 2.7 grams and maximum lift to weight ratio of 1.4 [1-3]. The design is based on the concepts developed by Campolo *et al.* [4, 5], which allows for direct control over the wing flapping angle and utilizes an elastic element for resonant operation as shown in Fig. 1. A flexure joint at the base of the wing allows for passive rotation driven by aerodynamic forces. Wings used in these studies were fully rigid due to modeling and fabrication constraints. In this study, we introduce a bio-inspired flexible wing design that is aimed at improving aerodynamic efficiency and lift production of such FWMAVs.

Biological wings are flexible structures that undergo inertial and aeroelastic deformation throughout the wing stroke [6, 7]. In general, flexibility is thought to increase the efficiency of flapping flight and provides high control authority based on modulation of wing kinematics [8-12]. Flexible wings have also been utilized by some successful FWMAVs, such as the Delfly and the Nano Hummingbird [13, 14]. The Delfly uses flexibility to achieve clap and fling motion that augments lift by 6%. Deformation of the wing was also observed to improve aerodynamic efficiency by 10%, although no direct mechanism for this result was proposed [15]. On the other hand, the Nano Hummingbird actively modulates wing flexibility for control. Although this system adjusts wing twist to control vehicle roll and pitch, the effect of twist on aerodynamic performance was not quantified. In addition to difficulties in quantifying flexibility and its aerodynamic consequences, just fabricating effective flexible wings can be a challenge. Our original version of this work, presented as a conference paper, explored the effects of wing flexibility more generally [16]. Wings with low overall stiffness were found to deform excessively, resulting in decreased effective wing area and over-rotation, which reduced the aerodynamic performance. Similar results are shown by Tanaka *et al.*, where chordwise deformations of a flexible polymer wing reduced performance compared to a rigid wing [17].

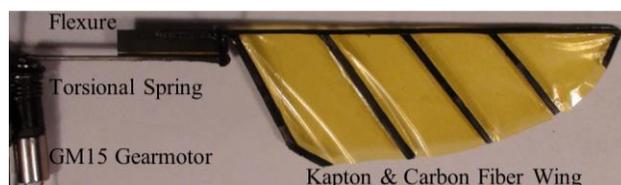

**Figure 1.** Picture of half our motor driven FWMAV. A torsional spring in parallel with the output of a DC gearmotor allows for resonant actuation. The wing, made of Kapton and carbon fiber, rotates passively about the flexure at its base.

The focus of this letter is to explore the flexibility effect of wing twist specifically, and to quantify its effect on translational lift production, the primary lift force generated by FWMAVs. This work is an updated version of our conference paper, where here we focus specifically on wing twist and further improve corresponding experimental and simulation results. Translational lift ($L_{trans}$) is modeled as:

$$L_{trans} = \frac{1}{2}\rho U(r)^2 c(r) C_l\big(\alpha(r)\big) dr, \quad (1)$$

where $\rho$ is the air density, $U(r)$ is the linear velocity at spanwise position $r$, $c(r)$ is the chord length, $\alpha(r)$ is the local angle of attack, and $C_l(\alpha)$ is the coefficient of lift at the given angle of attack, which is dependent on wing shape and Reynolds number (Re). Distance $r$ and differential element $dr$ are shown in Fig. 4b. This equation is based on a quasi-static blade element force model. Forces are calculated for chordwise blade elements and integrated along the length of the wing. Translational drag has the same form as Eq. 1, but uses the drag coefficient $C_d$. Force coefficients were determined experimentally for a scaled up model of *Drosophila* at Re ≈ 100 by Dickinson *et al.* shown in Fig. 2 [18]. Although our tests are performed around Re ≈ 10,000, previous results have found that the low Re results from Dickinson *et al.* show good agreement with inviscid models that are applicable to our Re regime [19-21].

Following, we will provide details on flexible wing design and fabrication in Section II. The experimental setup for wing characterization is described in Section III. Results are presented and discussed in Section IV. A conclusion on the effects of wing twist is presented in Section V.

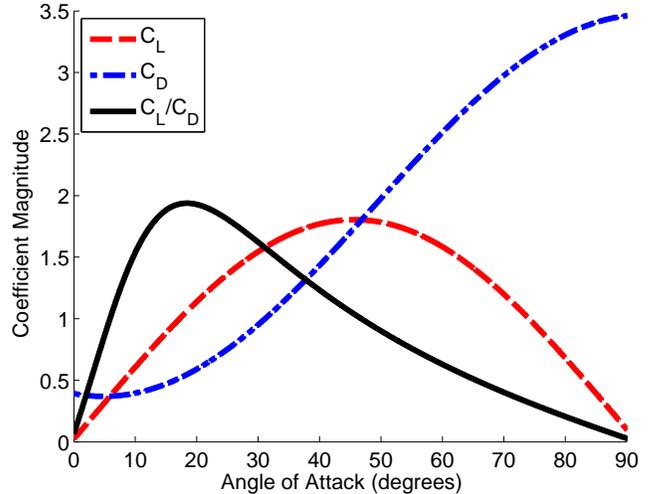

**Figure 2.** Lift and drag coefficients ($C_L$ and $C_D$) determined from experiments by Dickinson *et al.* [18] as a function of angle of attack. At 90° angle of attack, the wing is perpendicular to the airflow producing no lift and maximal drag. At 0°, the wing is parallel to the airflow.

## 2. Methods

Twisting of an airfoil changes the local angle of attack of individual wing sections as seen in Fig. 3. The blades of rotorcrafts are typically designed with a twist profile in order to control the lift distribution generated along the blade span. This is done because the incoming air velocity increases along the blade due to its radial motion. The twist of the airfoil is designed such that the resulting lift distribution is parabolic, which minimizes induced drag, primarily by reducing the angle of attack near the blade tip, where the air velocity is the highest [22]. Since our flapping wing has a similar velocity profile along its length due to its radial motion, generating a twist profile could also be beneficial for our FWMAV. Furthermore, wing twisting has been observed in many flying insects as well as hummingbirds [23-28]. Our original design was chosen to operate at a 45° angle of attack to maximize the lift coefficient. However, the lift to drag ratio for this operating point is close to unity as can be seen in Fig. 2. Decreasing the angle of attack towards the tip should increase the lift to drag ratio of these segments and improve wing efficiency. Although this would also decrease the overall lift coefficient of the wing, a more efficient system could achieve higher total lift with less power by increasing the flapping amplitude or frequency. It is also possible that the twisted wing shape could generate new aerodynamic behavior, resulting in modified force coefficients that improve overall aerodynamic efficiency and lift production.

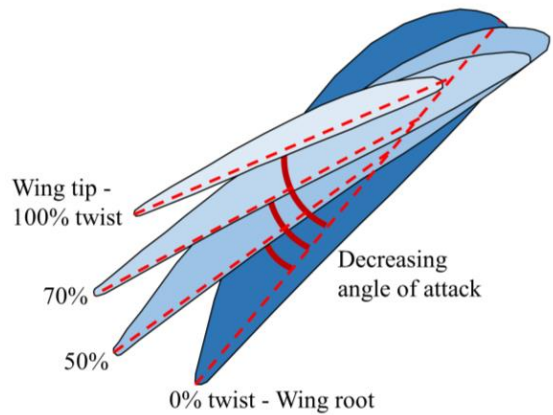

**Figure 3.** Four cross-sections of a twisted wing, each labeled with its percentage of the twist profile and colored by its position, darkest at the root. The largest section is at the base, with no twist. The next section is half the distance to the tip, with 50% of the applied twist profile. Proceeding out, the blade tapers and the chord line (shown as dashed line) rotates decreasing the angle of attack of the cross-sections, which is smallest at the tip.

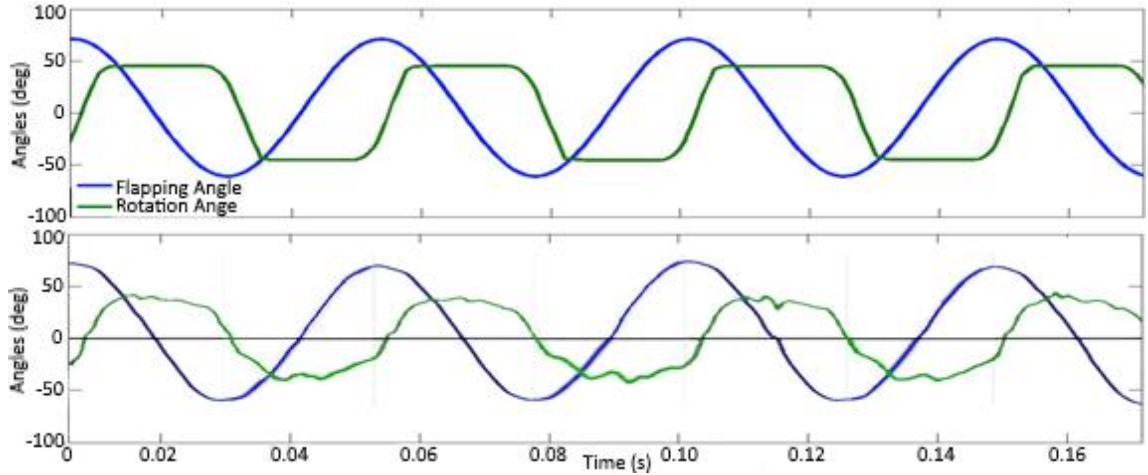

**Figure 5.** Experimental setup with control computer, power electronics, half system, load cell, and camera. The motor control signal is shown in black, while measured data is shown in green.

Fabricated wings are shown in Fig. 4. The wing structure was fabricated out of a unidirectional carbon fiber prepreg (M60J with 60% Toray 250° F Epoxy Resin) with a thickness of 40 μm and a 6-μm thick Kapton film as the wing membrane. The wing was laid up as a composite with Kapton held between structural carbon fiber layers. This assembly was then cured at 180 °C for two hours in a vacuum chamber. A carbon fiber rod was added to the leading edge to stiffen the wing. The assembly was completed with the addition of the flexure, rotational stoppers at ±45 degrees, and additional rod connecting back to the gearmotor shaft as shown in Fig. 1. Mathematically formalized wing shapes were described in work by Roll *et al.* and Ellington [29, 30]. Our chosen wing shape has a non-dimensional second moment of area ($\hat{r}_2$) of approximately 0.54 as shown in Fig. 4a. The length of the wing is 70 mm, with a mean chord of 20.5 mm, and a 38 mm offset is used from the wing base to the motor shaft. The rigid-wing design used two reinforcing spars and a 1 mm diameter carbon fiber rod was added to the leading edge to ensure the wing was fully rigid. For the flexible wing designs, Twist v1 and Twist v2, torsional stiffness had to be decreased such that the aerodynamic forces could dynamically shape the wing. Twist v1 used only half of a 1 mm rod along its length, while Twist v2 had half of a 1 mm rod for the basal 57% of wing with a 0.3 mm diameter rod supporting the rest of the leading edge to the tip. Furthermore, these designs used five supporting spars in order to provide additional locations where the distributed aerodynamic force on the wing membrane could be transmitted as a torsional load to the leading edge.

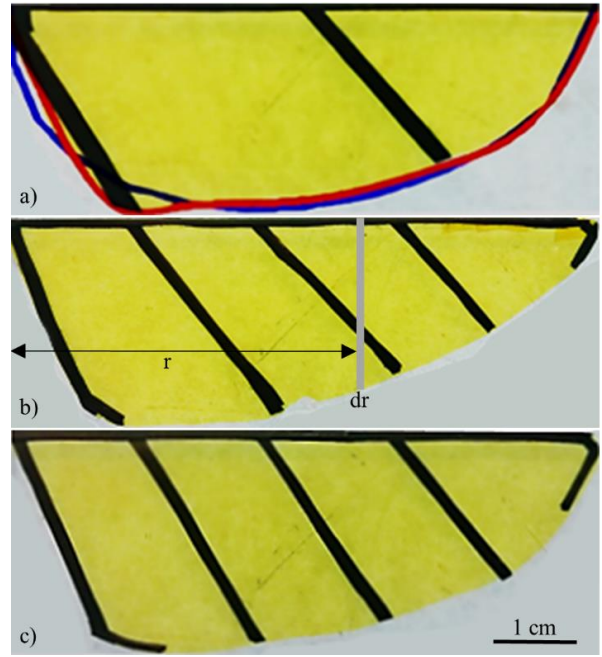

**Figure 4.** (a) Rigid wing with $\hat{r}_2 = 0.54$ beta distribution shown in blue and the modeled wing spline shown in red, and flexible wing designs, Twist v1 (b) and Twist v2 (c), with distance *r* and element *dr* shown.

Wings were tested using a control computer with two DAQ boards (National Instruments PCIe-6353 and PCI-6952e) as shown in Fig. 5. One card generated sinusoidal control signals for the motor driver (Dimension Engineering SyRen 10) that powered the motor. To determine the system resonance, the control signal frequency was varied from 10 to 30 Hz at constant input voltage amplitude of 4.9 $V_{rms}$ for each wing. The resonant operating frequency was selected based on minimum current draw, which also corresponds with the maximum flapping amplitude. The determined frequencies were 21, 22, and 23 Hz for the original, Twist v1, and Twist v2 designs, respectively, with variation due to differences in weight between designs. At the resonant frequency, each wing was tested at four input voltages: 4.9, 5.5, 6.2, and 7.2 $V_{rms}$. Lift was measured directly from the load cell (ATI Nano17 Titanium) by averaging wing strokes 5 to 125. The raw force data contained contributions from aerodynamic forces and inertial effects due to wing motion, particularly vertical motion during stroke reversal. The data was processed to remove inertial effects. Although this processing changed the magnitude of the measured lift force, the relations between wings remained similar and will be references throughout. Given that the wings were designed with consistent axes of rotation and symmetric rotational dynamics were measured from high-speed video, the contribution of rotational effects on stroke-averaged lift was expected to be negligible. Therefore the stroke-averaged forces were considered to be translational lift only. The voltage (measured directly as an analog signal) and current (ASC712-30A) output of the driver were measured by the second DAQ board and were averaged over the same wing strokes to calculate the input power to the motor. Inertial power was estimated from separate proof mass tests and subtracted from the measured power in order to calculate the aerodynamic power. Flapping amplitude was measured using Matlab image processing of high-speed video (PCO Dimax) images over the same wing strokes. Twist was also characterized from high-speed video images at mid-stroke, where the twist profile was fully developed by measuring the projected distance between points on the trailing edge and the leading edge. The difference between the projected distance measured from the video images and that of the untwisted wing was then used to calculate the local area of attack at each measured point.

## 3. Results and Discussion

Lift results as a function of the aerodynamic power input are shown in Fig. 6a. The tested powers corresponds with the nominal operating range of the rigid-wing during hovering of the vehicle, with the lowest point surpassing the takeoff lift requirement of 26 mN. Twist v1 and Twist v2 produced improvements of 41.3% and 22.1%, respectively (44.3% and 17.1% from raw data), over the rigid-wing design with respect to aerodynamic efficiency. These designs also increased translational lift production by 35.3% and 17.7%, respectively (38.1% and 12.9% from raw data), indicating that the twisted shape resulted in increased lift coefficients. Results of the calculated wing twist profiles are shown in Fig. 6b. The rigid-wing design displayed minimal twist as expected. Twist v1 was highly twisted with a 0.29 degree/mm profile for the basal 48 mm of the wing and a 1.39 degree/mm profile for the remaining 22 mm. Twist v2 displayed minimal twist for the basal 20 mm and had a constant 0.46 degree/mm profile for the remaining 50 mm of the wing.

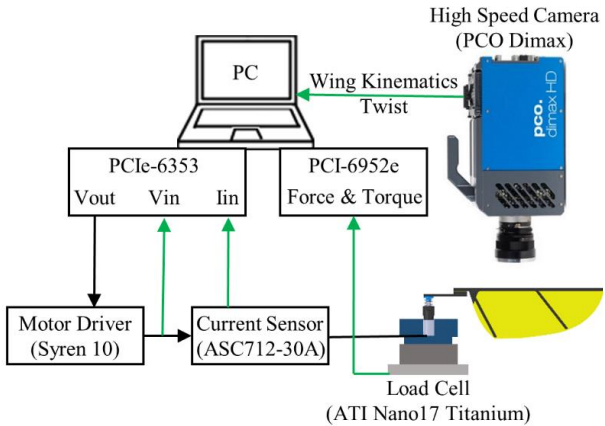

**Figure 5.** Experimental setup with control computer, power electronics, half system, load cell, and camera. The motor control signal is shown in black, while measured data is shown in green.

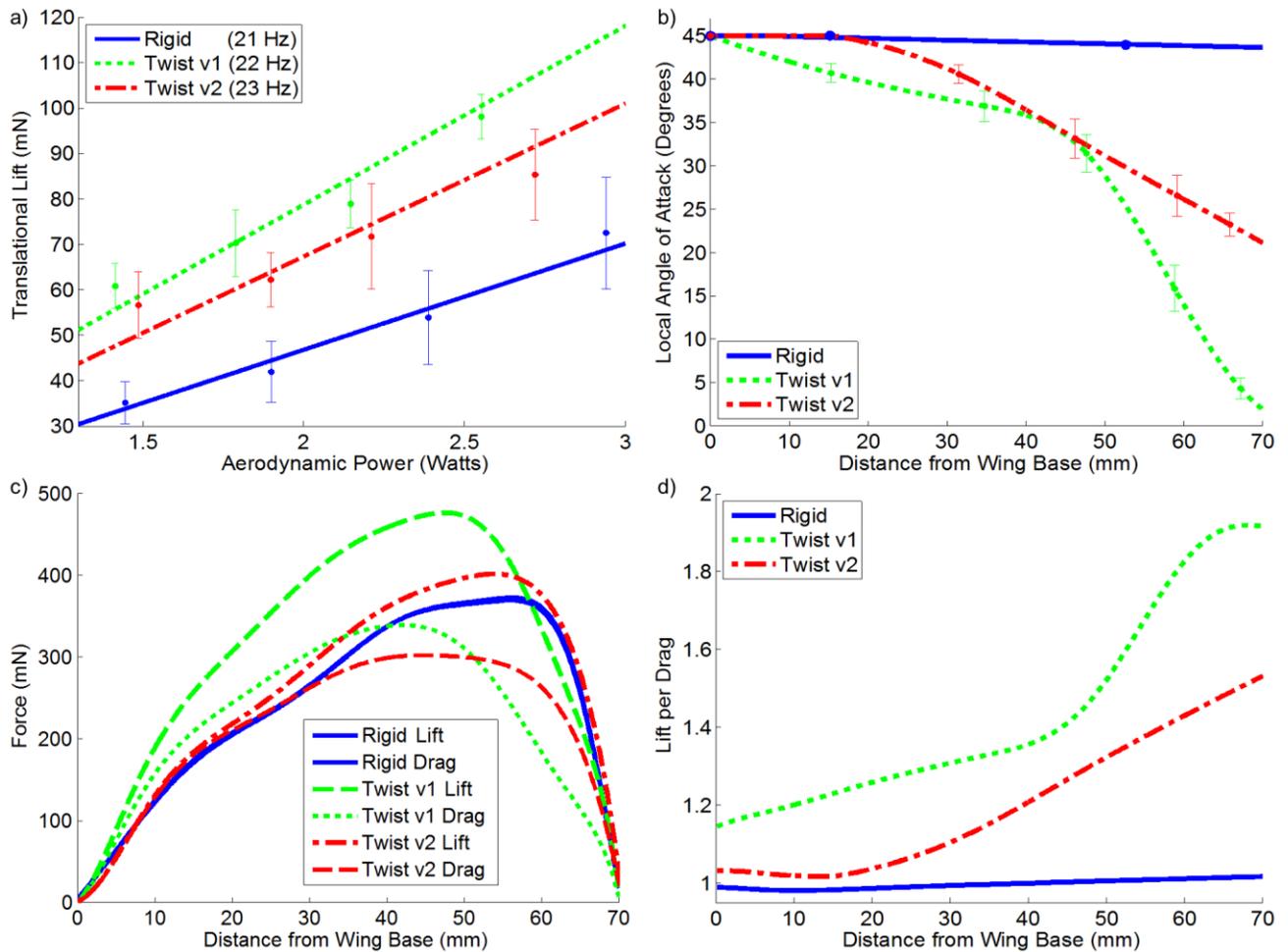

**Figure 6.** (a) Experimental lift results as a function of input aerodynamic power. (b) Calculated twist profiles from experimental high-speed video images. (c) Simulated lift and drag distributions based on twist profiles. Lift and drag are equal for the rigid-wing design, and appear as a single line. (d) Simulated lift per drag, aerodynamic efficiency, along the wing length based on twist profile.

These twist profiles were then used to model the stroke-averaged translational lift and drag forces along the wing as shown in Fig. 6c. The instantaneous half-stroke wing velocity field was modeled based on the wing flapping amplitude, wing rotation, and operating frequency. Dynamic twisting of the wing was included based on high-speed video analysis such that the wing linearly transitioned from flat to fully twisted over the first 20% of the stroke, remained fully twisted for the subsequent 60% of the stroke, and returned to flat over the final 20%. A quasi-steady blade element model was used to estimate the lift and drag forces per segment based on the force coefficients from Fig. 2, which were then integrated along the wingspan. To estimate changes in force coefficients, the total lift was scaled to match the experimental results. Aerodynamic power was also scaled by the experimental values and then normalized such that the rigid-wing operated at a constant lift per drag ($L/D$) of one. The resulting $L/D$ along the wings is shown in Fig. 4d. This analysis shows increases in the effective lift coefficient by 63.6% and 13.3%, respectively (67.1% and 8.6% from raw data), for Twist v1 and Twist v2. It is possible that the twisted shape of the wing stabilizes or strengthens the leading edge vortex (LEV) resulting in observed increase in the lift coefficient. This could occur directly due to the 3D shape of wing, increased spanwise flow, or decreased angle of attack near the tip could prevent bursting of the LEV. Such mechanisms for strengthening of the LEV are discussed in work by Lentink *et al.* [31]. The trend of increasing aerodynamic efficiency and lift production is also consistent with computational fluid dynamics (CFD) results by Noda *et al.* for flexible wings [32]. Twist v2 displays a moderate twist profile that improves efficiency along the entire wingspan, primarily through a reduction in drag. Although lift production is slightly improved, the lift distribution remains similar to the rigid-wing design that is biased towards the wing tip. The highly twisted profile of the Twist v1 design greatly improved efficiency along the wing, especially near the tip. The lift and drag distributions are re-shaped to better match the ideal elliptical distribution. This design produces significantly more lift in the center of the wing, while achieving similar drag to the other designs. Lift towards the tip is similar to the other designs, but with significantly reduced drag likely due to a reduction of wing tip vorticity.

Now that the performance of our fabricated wings has been characterized, it can be compared to optimal designs predicted by the quasi-steady model. Integrating under the lift curve in Fig. 6c gives total stroke-averaged lift, while integrating under the *L/D* curve in Fig. 6d provides a metric of flapping efficiency. These lift and efficiency metrics were co-optimized using MATLAB ga to generate a Pareto frontier that shows all potentially optimal twist profiles for different weightings of lift versus efficiency. The optimization was constrained to strictly increasing twist profiles that take advantage of the increasing oncoming flow velocity along the wing to generate twist. The frontier demonstrates the expected tradeoff between lift production and efficiency. Wing twist, which decreased the average angle of attack, results in a more efficient wing that produces less lift. Furthermore, the simulation results indicate that this tradeoff can be achieved by uniformly changing the angle of attack of the entire wing, effectively adjusting the rotational stoppers for a rigid wing. This could be implemented with the use of additional actuators as is done by the Nano Hummingbird, but ultimately reduces the lift per weight of the vehicle [14]. Our experimental results indicate that the standard quasi-steady model does not account for changes in force coefficients due to 3D twisted wing shapes. Therefore, the simulation was updated to include a 3D shape factor, fit from experimental data, which accounts for the lift augmentation due to increasing wing twist profiles.

The new Pareto frontier is shown in Fig. 7. This frontier indicates a range of optimal average wing twists between 1.28 and 1.45 lift per drag. The resulting twist profiles are shown in Fig. 8. Wings producing the most lift have low twist in the basal half of the wing with uniformly increasing twist along the distal portion, similar to the Twist v2 design. However, the lift production compared to Twist v2 is increased by 9.5% by increasing the basal twist from an average of 0.17 degrees/mm to 0.27 and distal twist from 0.51 degrees/mm to 0.83. Wings designed for higher efficiency display a two-stage twist profile similar to Twist v1, although the model falls short of predicting the lift achieved by this design. However, there are some differences in the shapes of these designs. Twist v1 has a nearly uniform twist profile of 0.25 degrees/mm for the basal 45 mm of the wing, whereas the model designs utilize 0.83 degrees/mm for the basal 26 mm of the wing followed by an untwisted section. The distal profile in Twist v1 is 1.27 degrees/mm, while it is 0.79 in the model. The importance of high twist near the wing tip agrees with CFD results by Noda *et al.* [32]. Improved CFD modeling could improve the understanding of the relationships between 3D wing shapes resulting changes in force coefficients, which could be applied to generate better estimates of optimal twisted designs. Furthermore, our results indicate that wing designs with significant twist profiles can provide high performance without active

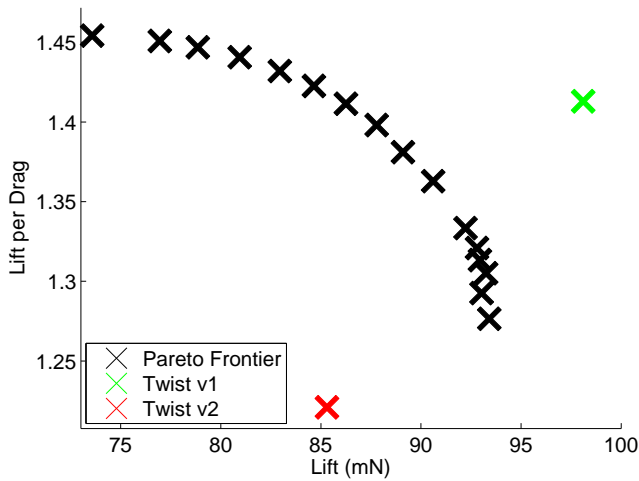

Figure 7. Pareto frontier of simulated optimal twisted wing designs.

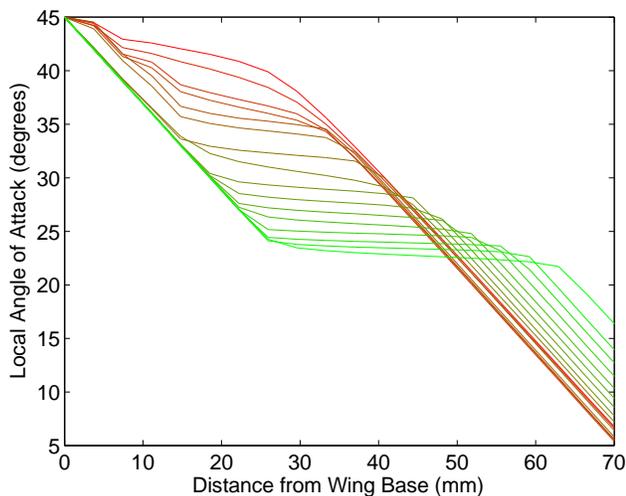

Figure 8. Simulated optimal wing twist profiles. Curves go from high lift (top, red) to high efficiency (bottom, green).

changes in stopper position, which is essential for low mass, underactuated FWMAV systems.

## 4. Conclusion

In summary, we have shown that a bio-inspired twisted wing design increased aerodynamic efficiency, translation lift, and the effective lift coefficient by 41.3%, 35.3%, and 63.7%, respectively, compared to our rigid-wing design. A quasi-steady blade element model predicted that twisting of the wing could improve efficiency, but at the cost of decreased lift production. This discrepancy between the model and experimental results indicated that the full 3D twisted shape of the wing changed the lift and drag coefficients through unsteady mechanisms, such as LEV augmentation. The quasi-steady model was updated with a 3D shape factor to estimate the lift improvement due to wing twist. The improved simulation was used to optimize twisted wing shapes, resulting in a series of twist profiles similar to the fabricated wings. However, the optimization fell short of predicting the lift achieved in experiments by the highly twisted Twist v1 design. Improved force coefficient estimation for flexible wings remains as an open problem. However, most studies focus on overall flexibility and fluid structure interaction, instead of on specific flexibility results, such as dynamic twisting or cambering. Future work will address wing twisting-based LEV augmentation with particle image velocimetry experiments. Furthermore, improved CFD modeling may provide direct relationships between twist profiles and resulting changes in force coefficients. This would allow vehicle designers to better optimize twisted wing shapes and would serve as a general design principle by which to improve the performance of bio-inspired FWMAVs.

## Acknowledgements

This work is supported by the National Science Foundation Graduate Research Fellowship Program under Grant No: DGE-1252522.